# Electronic disorder in crystalline organic semiconductor: the role of thermal motions modulating the transfer integrals in polyacenes

Alessandro Troisi and Giorgio Orlandi

*Dipartimento di Chimica 'G.Ciamician' Università di Bologna,via F. Selmi 2, 40126 Bologna, Italy*

*Abstract*

The effect of thermal structural fluctuations on the modulation of the transfer integrals between close molecules is studied using a combination of molecular dynamics simulations and quantum chemical calculations of the transfer integral. We have found that the fluctuations of the transfer integrals are of the same order of magnitude of their average value for two common organic semiconductors an observation that puts into question the currently adopted models for charge transport in organic semiconductors and identifies in the dynamic electronic disorder the limiting factor for the charge carrier mobility.

*E-mail: a.troisi@warwick.ac.uk

The theoretical understanding of the charge transport mechanism in ordered organic semiconductors is still unsatisfying. Notwithstanding decades of investigation,[1-3] several contradictions between known conduction models and experimental evidences suggest that the framework developed for the study on inorganic semiconductors is somewhat inadequate for the rationalization of the transport properties of organic materials.[4] For example, in pure thin film of pentacene the hole mobility $\mu$ decreases as the temperature increases up to room temperature according to a power law ($\mu \sim T^n$),[5] a phenomenology compatible with the band transport mechanism. However, the fitting of the experimental data with the phenomenological models indicate a mean free path of the hole comparable with the crystal unit cell,[6, 7] a fact in contradiction with the delocalized picture implied by the band model.

It is convenient to recall the structural and electronic differences between inorganic and organic semiconductors to focus on the origin of the different transport mechanisms. Crystalline organic semiconductors are narrow band materials, i.e. the electronic coupling between the molecules in the crystal is much weaker than the electronic coupling among atoms in silicon or germanium. On the other hand, the unit cell in organic materials is much larger causing an effective mass for the charge carrier comparable to that of wide band materials.[8] The dispersive interactions between the molecular constituents of organic semiconductors are weak, making these materials much softer than their inorganic counterpart with consequences that affect also the transport mechanism. This point is very well exemplified by several studies in the related study field of charge *transfer* in donor-bridge-acceptor systems. If the 'bridge' is made by *covalently bonded atoms* (as in silicon) the coupling between donor and acceptor can be considered constant (i.e. Condon approximation holds).[9] When the transfer takes place through *non bonded molecules* (as in organic crystal), thermal motions of the nuclei cause a fluctuation of the donor-acceptor coupling that affects the charge transfer rate.[10] Examples of non covalently bonded bridges that produce a strong fluctuation of the intermolecular coupling include, among others, the DNA base pair,[11] some proteins,[12] or the solvent[13] in specially designed systems. The experience from charge transfer studies suggests that the electronic coupling between (non bonded) molecules also in an organic semiconductor can be strongly influenced by the thermal structural fluctuations.

In this paper we assess the importance nuclear thermal motions in the modulation of the intermolecular coupling relevant for the charge transport in organic semiconductors. We present the results for the pentacene, probably the most studied compound for organic electronics forming ordered crystal or thin film structures. Results relative to the anthracene crystal will be also presented for comparison.

**Model**. The full quantum description of the electron-phonon coupling in an organic semiconductor is complicated and it usually requires some assumption on the coupling strengths that allows some sort of perturbative treatment.[14, 15] In this paper we adopt a semiclassical approach assuming that the nuclear motions of the crystal can be treated classically while the charge carriers are treated quantum mechanically, an assumption often taken in the study of conductive polymers. We use the following time-dependent one-particle (hole or electron) Hamiltonian:

$$H = \sum_i \varepsilon_i a_i^+ a_i + \sum_{i \neq j} V_{ij}(t) a_i^+ a_j, \qquad (1)$$

where the operators $a_i^+$ and $a_i$ create and annihilate a fermion in a molecular orbital (MO) localized on the *i*-th molecule. We assume one relevant MO per molecule, for example the HOMO orbital if we are interested in the hole transport. The time dependence of $V_{ij}(t)$ is imposed by the classical nuclear motions that (in this model) have the only effect of modulating the transfer integral. The approximations implied in this approach and some justifications are summarized below:

(i) A single band can be used to describe the electron/hole. Band calculations show that the bands originating from the HOMO and the LUMO do not mix with bands originating with other orbitals in most organic semiconductors including pentacene and that, therefore, a single band provides a valid description of the electron or hole mobility.

(ii) The phonons that couple with the electrons are low frequency modes that can be treated classically. This point can and will be verified *a posteriori*. The semiclassical approach has the advantage that the effect of tens of optical phonon modes, needed to describe the nuclear motions of an organic crystal, is summed up in a unique time dependent quantity $V_{ij}(t)$.

(iii) The classical motion of the nuclei is marginally affected by the charge carrier wavefunction and $V_{ij}(t)$ can be seen as an external modulation. An extra positive (or negative) charge on a large molecule like pentacene causes minor changes in its molecular vibrational modes[16] suggesting that the modulation of the transfer integral $V_{ij}$ is not affected by the charge carrier distribution. The principal finding of this paper is not affected by this approximation that can be relaxed at a later stage.

(iv) We are neglecting the diagonal or Holstein coupling, i.e. the coupling of the matrix elements $\varepsilon_i$ with the nuclear coordinates. These terms account for the deformation of the molecule when a net charge is on site *i*. The effect of Holstein coupling is extensively discussed in literature and textbooks.[17] Its main effect, in the small coupling regime appropriate to pentacene, is to produce band narrowing[14] with consequent reduction of the charge mobility (while in the large coupling regime it is responsible of the formation of small polarons). This approximation is introduced only for simplicity because we will focus in this paper on the importance of off-diagonal electron phonon coupling.

The main system investigated in this paper is the (001) plane of crystalline pentacene, one of the reference standards for organic electronics. An ordered pentacene thin film can be grown on a $SiO_2$ substrate[18-21] with the *ab* plane parallel to the substrate. Hole conduction takes place mostly within this plane (probably the first layer)[22] in a thin-film transistor experimental set-up. The highest occupied band of the perfectly ordered (001) plane of pentacene can be expressed analytically as a function of three transfer integrals between the (umperturbed) HOMOs of the isolated molecule.[23] We investigate here the effect of thermal motions on these three transfer integrals.

**Computational Details**. We replicated the crystal unit cell[24] building a 3x2x2 supercell (containing two layers of pentacene molecules in the *ab* plane). The dynamics of the system was studied with periodic boundary conditions employing the MM3 force field,[25] that allows a correct description of the bond alternation in conjugated compounds. We run MD simulations at fixed volume and constant temperature[26] of 100K, 200K, 300K, using an integration time step of 2fs (and hydrogen atoms kept at their equilibrium distance[27]).

We evaluated the transfer integral between HOMO orbitals every 30 fs with the INDO/S Hamiltonian, which proved to give results in reasonable agreement with more accurate DFT calculation.[28] As reported in Figure 1a, the close molecular pairs can be of type *A*, *A'*, *B* and *C*. For each considered snapshot we computed 3 transfer integrals between couples of type *A* and *A'*, and 6 transfer integrals between couples of type *B* and *C*. The 18 evaluations of the intermolecular coupling were the slowest part of the procedure that limited the length of the analyzed dynamics to 100ps. We note that, in the crystal plane of interest, no molecule interacts with its image. The very same procedure was applied to the *ab* plane of the crystalline anthracene,[29] displaying a packing similar to that of pentacene. This additional computation was done to verify that the characteristics shown in this paper are likely to be general features of polyacenes and, probably, of all crystalline organic semiconductors.

**Results**. We will begin the analysis describing the distribution of the HOMO-HOMO coupling between molecular pair and neglecting its time dependence. Since the couplings *A* and *A'* behave similarly, we will mention only the former in the discussion. Table 1 shows the average couplings and the variance computed for the pentacene and the anthracene at different temperature. Couplings of the same type between different molecules are treated cumulatively in the calculation of the averages. Figure 1b displays the distribution of the *A*, *B* and *C* coupling for the pentacene at different temperatures. The distributions are Gaussian with width varying between 250 and 400 cm$^{-1}$ at room temperature.

The variance of the transfer integrals coupling is extremely large and the modulation introduced by the nuclei is of the same order of magnitude of the average coupling. To search for possible correlations between the intermolecular coupling $V_K$ and $V_L$ we computed the statistical correlation between them, defined as:

$$\mathrm{cor}(V_K, V_L) = \frac{\langle V_K V_L \rangle - \langle V_K \rangle \langle V_L \rangle}{\sigma_{V_K} \sigma_{V_L}}, \qquad (2)$$

being $\sigma_{V_K}$ and $\sigma_{V_L}$ the standard deviation of the individual coupling. The correlation (computed between all the couples shown in Fig.1 at 300K) is small ($0.1 < |\mathrm{cor}(V_K, V_L)| < 0.25$) for pairs of intermolecular couplings that share a common molecule and negligible ($|\mathrm{cor}(V_K, V_L)| < 0.1$) for independent couples of molecules. This observation can be very helpful for the modeling of the process because, to a very good degree of approximation, the intermolecular coupling fluctuation can be considered uncorrelated. The lack of correlation is due to the relatively large number of phonon modes that are coupled to the intermolecular transfer integral. Many low frequency optical phonons are essentially dispersionless, i.e. they can be also treated as vibration localized on each molecule (as in the Einstein model). These modes contribute with independent phases to the time dependence of *V*(*t*) and the overall coupling pattern is essentially random.

The effect of uncorrelated modulation is equivalent to the effect of *disorder* on the electronic structure and its main effect is the *localization* of the hole wavefunction.[30-32] The localization of the electron eigenfunctions (in the frozen conformation at a give time) of a system characterized by the parameters given in Table 1 can be easily evaluated. We built a large supercell containing 3200 pentacene molecules, coupled according to the pattern outlined in Fig. 1a, and we attributed random values to the transfer integrals following a Gaussian distribution with the parameters of Table 1 (i.e. derived from the computations on a small supercell). A quantitative measure of the degree of localization of an eigenfunction can be given by the number of molecules with the highest charge density that account for the 50% of the total (unit) excess charge, defined as $N_{50\%}$. The localization decreases as the energy of the eigenfunctions increases. The Boltzmann average of $N_{50\%}$ is 39, 36 and 31 molecules at 100, 200 and 300K respectively. At low temperature the disorder and the consequent localization is less pronounced but higher energy and more delocalized states are less populated. The combination of these two factors causes a relatively modest effect of the temperature on the degree of localization of the wavefunction.

A convenient way to evaluate the role of the coupling dynamics is plotting the Fourier transform of the autocorrelation function $\langle \delta V_K(0) \delta V_K(t) \rangle$,[10] shown in Figure 2 for the couplings $A$, $B$, $C$ (we have defined the deviation from the average transfer integral $\delta V_K(t) = V_K(t) - \langle V_K(t) \rangle$). The most important point is that the majority of the phonon modes that modulate the coupling have energy around 40 cm$^{-1}$ and that the contribution of modes above 160 cm$^{-1}$ is negligible. Since only low frequency modes play a role in the modulation of the transfer integral, the semiclassical description adopted here can be considered valid at room temperature and at least plausible above 150K (vibrations can be treated classically if their frequency ω is such that $\hbar\omega \ll k_BT$). The other important point is that the dynamic disorder of organic crystals has a very long correlation time: $V_K(0)$ and $V_K(t)$ are correlated for time longer than the MD simulation length (in such cases the correlation time is not precisely determinable). The long correlation time distinguishes these systems from conductive liquid or liquid crystals.[33, 34] There is, as may be expected, some overlap between the peaks in Fig. 2 and the Raman spectra of the pentacene,[35] a resemblance that we take as a confirmation of the reliability of the used force field.

**Discussion**. The computational results presented here strongly points to an essential deficiency in the theoretical modeling of charge transport in organic material: *it is not possible to treat the electron-phonon coupling as a perturbation*. The modulation of the transfer integrals, due to low frequency phonons, is of the same order of magnitude of their average values already above 100K. The band description of the material breaks down since the *k* vector is not a good quantum number for systems with this degree of disorder. Thermal disorder of the electronic Hamiltonian is able to localize the charge carrier within few unit cells preventing a description of the transport based on the carrier mean free path, which is consistent only with a delocalization of the carrier over a distance of many unit vectors. A similar behavior has been discussed for several metals, known as "bad metals", and is related to the approaching of the electrons mean free path to the unit cell length.[36, 37]

The huge overall modulation of the transfer integral $V_{ij}$ is due to the high number of low frequency modes that contribute independently to it (see Fig.3). In a typical perturbative approach the effect of each phonon mode is treated independently through the electron phonon coupling matrix elements taking the form of $\partial V_{ij}/\partial u_\lambda$ ($u_\lambda$ is the nuclear displacement along the phonon mode λ).[14] While each coupling term can be relatively small the global effect is dramatic if the unit cell contains tens of modes for which the coupling is not negligible (an essential difference between organic and inorganic semiconductors). The phonons in such systems affect the zero-th order description of the electronic wavefunction making the theoretical modeling of the transport particularly challenging. The disorder of these materials is *dynamic* and a transport model should be able to describe the time evolution of localized states such as the ones depicted in Figure 2. The calculation presented in this paper suggests that a semiclassical numerical approach can be used to study the evolution of the charge carrier wavefunction in a model system characterized by few parameters characterizing its dynamical disorder like the average amplitude of the modulation and the average frequency of the modulating phonons.

In conclusion, computational chemistry methods have been used to evaluate the role of thermal motions in the modulation of the transfer integrals between close molecules. We have found that the fluctuations of the transfer integrals are of the same order of magnitude of their average value for pentacene and anthracene an observation that puts into question the currently adopted models for charge transport in organic semiconductors and identifies in the dynamic electronic disorder the limiting factor for the charge carrier mobility. The present results will contribute to the definition of more realistic transport models for organic crystals.

**Table 1**. Average coupling (cm$^{-1}$) and variance of the intermolecular transfer integral between molecular HOMOs in pentacene (space group $P\bar{1}$) and anthracene (space group P2$_1$/a) at different temperatures. The higher symmetry of the anthracene crystal makes equivalent the transfer integrals A-A' and B-C.

|            | T=100K   |            | T=200K   |            | T=300K   |            |
|------------|----------|------------|----------|------------|----------|------------|
| Pentacene  | <V>      | $\sigma_V$ | <V>      | $\sigma_V$ | <V>      | $\sigma_V$ |
| A          | 459.8    | 150.3      | 460.9    | 199.8      | 455.3    | 258.8      |
| A'         | 477.2    | 150.5      | 485.5    | 207.1      | 476.6    | 262.1      |
| B          | -630.9   | 208.0      | -626.0   | 291.3      | -615.9   | 356.4      |
| C          | 1018.4   | 237.5      | 1008.0   | 327.1      | 983.3    | 404.4      |
| Anthracene |          |            |          |            |          |            |
| A          | -621.9   | 203.7      | -616.7   | 275.8      | -629.6   | 375.7      |
| B          | 445.7    | 276.3      | 439.8    | 395.1      | 426.7    | 496.6      |

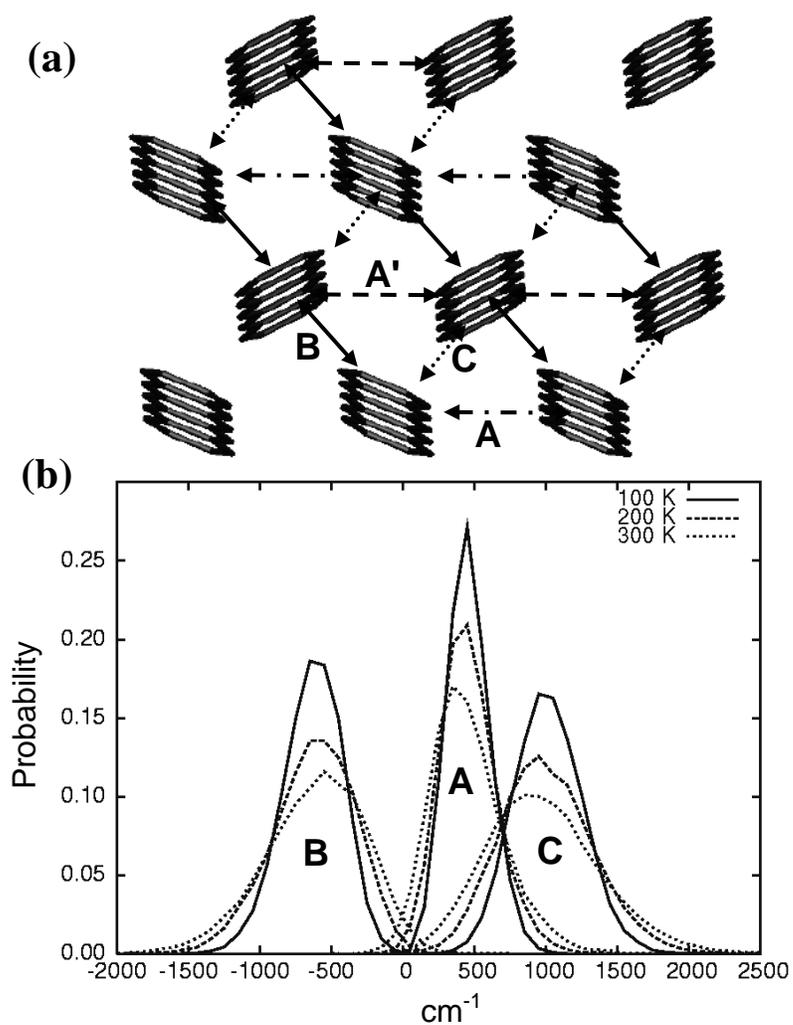

**Figure 1**. (a) Six unit cells in the *ab* plane of pentacene (a portion of the supercell considered for the MD simulation). The arrows indicate the intermolecular couplings monitored along the MD, divided in four groups *A* (dashed-dotted), *A'* (dashed), *B* (solid) and *C* (dotted). (b) Distribution of the transfer integrals of solid pentacene at different temperatures.

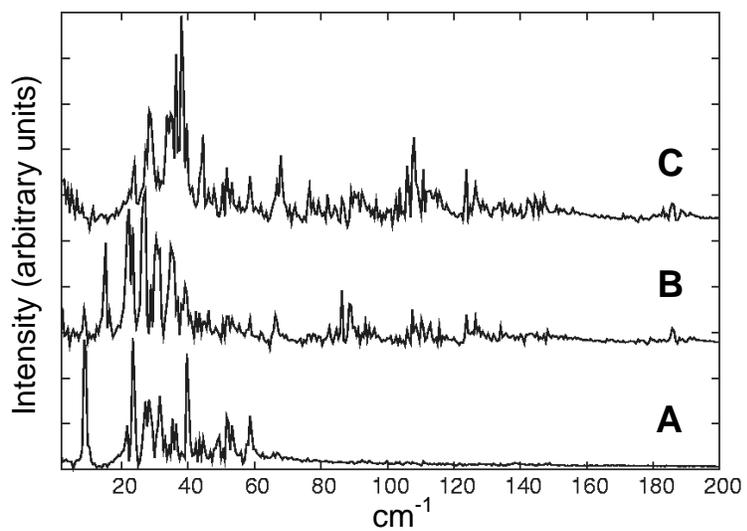

**Figure 2**. Fourier transform of the autocorrelation function of the transfer integrals between HOMOs of neighboring molecules (300K). A vertical offset was introduced for clarity among the curves relative the transfer integral type *A*, *B*, and *C*.